\documentstyle[aps
,preprint%
]{revtex}
\newcommand{\be}{\begin{equation}}
\newcommand{\ee}{\end{equation}}
\newcommand{\bea}{\begin{eqnarray}}
\newcommand{\eea}{\end{eqnarray}}
\newcommand{\nn}{\nonumber \\}
\newcommand{\Bx}{\mbox{\boldmath$x$}}

\newcommand{\Bv}{\mbox{\boldmath$v$}}
\newcommand{\BV}{\mbox{\boldmath$V$}}

\begin{document}
\draft
\preprint{AJC-HEP-29}

\date{\today
}

\title{
Low-Energy Interaction of a Cosmic String and
an Extreme Dilatonic Black Hole
}

\author{Kiyoshi~Shiraishi%
\thanks{e-mail: {\tt g00345@sinet.ad.jp,
shiraish@air.akita-u.ac.jp}
}}

\address{Akita Junior College\\
Shimokitade-sakura, Akita-shi, Akita 010, Japan
}

\maketitle

\begin{abstract}
The interaction of a cosmic string and a maximally charged dilatonic
black hole is studied in the low-velocity limit.
In particular, the string-black hole scattering at a low velocity
is investigated.
\end{abstract}

\vspace{7mm}
\vfill
\eject


Cosmic strings~\cite{VS} are topological defects,
which may have resulted from
a certain phase transition in the early universe.
When we take the infinitely thin limit of the string core,
the cosmic string is characterized by a single dimensionless
number $G\mu$, where $\mu$ is
the mass density per unit length (in the unit $c=1$)
and $G$ is the Newton constant.
The spacetime outside of a straight cosmic string is described
by the metric~\cite{Vil}
\be
ds^{2}= - dt^{2} + dr^{2} + \frac{r^{2}}{\nu^{2}} d\theta^{2}
+ dz^{2}\; ,
\label{eq:csmet0}
\ee
where $\nu=(1-4G\mu)^{-1}$.

If we use a new coordinate
\be
\vartheta=\frac{\theta}{\nu}\; ,
\ee
the metric (\ref{eq:csmet0}) reduces to
the one of the flat spacetime
\be
ds^{2}= - dt^{2} + dr^{2} + r^{2} d\vartheta^{2} + dz^{2}\; ,
\label{eq:csmet1}
\ee
but there is a deficit in the azimuthal angle since $\vartheta$
has the range
\be
0\leq\vartheta<\frac{2\pi}{\nu}\; .
\ee
Then the space has a conical singularity along $z$-axis in
the infinitely thin limit of the string.
The deficit angle $\Delta$ is defined as~\cite{VS,Vil}
\be
\Delta\equiv 2\pi\left(1-\frac{1}{\nu}\right)=8\pi G\mu\; .
\ee

Linet~\cite{Linet} and Smith~\cite{Smith} have shown that
a charged test particle placed near an
infinite straight cosmic string feels a static self-force caused by
the conical structure of the spacetime.
It is also interesting to examine how the conical structure
provides some kinematical effects in the motion of compact objects,
since
the interaction between cosmic strings and moving black holes may
frequently occur in the early stage of the universe.
In this paper, we construct the solitonic solution in
Einstein-Maxwell-dilaton theory on the conical spacetime and
investigate the motion of the solitonic object at a low velocity.

The reason why we adopt this particular model is that
the low-energy interaction between such dilatonic objects
has been well known and we can utilize the result about it in
the present case.
The exact solution for an arbitrary number of
charged dilatonic solitons has been found
by several authors~\cite{GHS,Shi1}.
In the following, using the method of images,
we obtain a solitonic solution on a conical space.
We find that the solution reduces to the extreme dilatonic black hole
solution on  $R^{4}$ if the deficit angle vanishes.%
\footnote{The multi-soliton system in
Einstein-Maxwell-dilaton theory on the spacetime
with torus compactification has been studied by
obtaining exact solutions and low energy interactions
in a similar method~\cite{Shi6}.}


We consider a model described by the following action:
\be
S=\int d^{4}x\ \frac{\sqrt{-g}}{16\pi G} \left[ R -
2(\nabla \phi)^{2} - e^{-2\phi} F^{2}
\right] +
\mbox{(surface terms)}\; .
\ee
where $R$ is the scalar curvature, $\phi$ a dilaton,
and $F_{\mu\nu}=\partial_{\mu}A_{\nu}-\partial_{\nu}A_{\mu}$
is the abelian gauge field strength.

We assume that the electric solution in the Einstein-Maxwell-dilaton
theory takes the form~\cite{GHS,Shi1}
\be
ds^{2}= - \frac{1}{V} dt^{2} + V d\Bx^{2},\;
A_{\mu}dx^{\mu} = \frac{1}{\sqrt{2}}\left(1-\frac{1}{V}\right)dt,\;
e^{-2\phi}=V\; .
\label{eq:met}
\ee
Then $V$ satisfies
\be
\nabla^{2}V=0\; ,
\label{eq:lap}
\ee
up to a number of delta functions
in the right hand side of Eq.~(\ref{eq:lap}).

For example, one can find a solution:
\be
V=1+\sum_{\alpha}
\frac{2 G m_{\alpha}}{|\Bx -\Bx_{\alpha}|}\; .
\label{eq:sol}
\ee
By studying the relation among
the electric and dilatonic charge and mass
of each solitonic object, we find that
the solution describes the configuration that the $\alpha$-th
nonrotating, charged dilatonic ``black hole''
in the extreme limit with mass $m_{\alpha}$
located at $\Bx =\Bx_{\alpha}$~\cite{GHS,Shi1}.
Strictly speaking, there are
singularities at $\Bx =\Bx_{\alpha}$.
Nevertheless, we use the term ``black hole'' because the extreme case
may still have generic properties of black holes in terms of
their classical dynamics. Thus we use the term ``black hole'' without
double quotation marks hereafter.


Now let us construct the black hole solution on a conical space.

When the string parameter $\nu$ is equal to an integer
$p=1,2,3,\ldots$, we can derive the solution for $V$, which
corresponds to the static configuration of
one extreme dilatonic black hole and a cosmic string
at the origin, by the method of images~\cite{Smith}.
The function $\Gamma_{p}(\Bx,\Bx')$ satisfying Eq.~(\ref{eq:lap})
and having periodicity $2\pi/p$ in $\vartheta$ and $\vartheta'$
can be expressed by the summation of $p$ images of $\Gamma(\Bx,\Bx')$,
which is a solution of  Eq.~(\ref{eq:lap}) in the Minkowski space:
\be
\Gamma_{p}(\Bx,\Bx')=\sum_{n=0}^{p-1}\Gamma(\Bx,\Lambda^{n}\Bx')
=\sum_{n=0}^{p-1}\frac{1}{|\Bx-\Lambda^{n}\Bx'|}\; ,
\ee
where $\Lambda$ means the rotation by $2\pi/p$ around the origin.
Using the integral representation, we then obtain
the solution for $V$ as follows:
\bea
\Gamma_{p}(\Bx,\Bx')&=&\sum_{n=0}^{p-1}
\frac{1}%
{\sqrt{r^{2}+{r'}^{2}-
2rr'\cos (\vartheta-\vartheta'-2\pi n/p)+(z-z')^{2}}}\nn
&=&\frac{1}{\pi\sqrt{2rr'}}
\int_{u_0}^{\infty}
\frac{du}{\sqrt{\cosh u - \cosh u_{0}}}
\sum_{n=0}^{p-1}
\frac{\sinh u}{\cosh u - \cos (\vartheta-\vartheta'-2\pi n/p)}\nn
&=&\frac{1}{\pi\sqrt{2rr'}}
\int_{u_0}^{\infty}
\frac{du}{\sqrt{\cosh u - \cosh u_{0}}}
\frac{p\sinh pu}{\cosh pu - \cos p(\vartheta-\vartheta')}\; ,
\eea
where
\be
\cosh u_{0}=\frac{r^{2}+{r'}^{2}+(z-z')^{2}}{2rr'}\; .
\ee

We have the solution of $V$ which corresponds to
asymptotically flat space:
\be
V=1+2 G M \Gamma_{p}(\Bx,\Bx')\; ,
\label{eq:solG}
\ee
where $M$ is a constant.
Note that $V$ is a solution of Eq.~(\ref{eq:lap})
even if $p$ is replaced by an arbitrary real number $\nu$.
Consequently, we have the solution of $V$ corresponding to
an extreme dilatonic black hole located at  $\Bx'=(r',\theta',z')$
in the conical space:
\be
V=V_{CS}=1+
\frac{2 G M}{\pi\sqrt{2rr'}}
\int_{u_0}^{\infty}
\frac{du}{\sqrt{\cosh u - \cosh u_{0}}}
\frac{\nu\sinh \nu u}{\cosh \nu u - \cos (\theta-\theta')}\; .
\label{eq:solcs}
\ee
This result can be extracted by using the mode expansion
in terms of the Legendre functions~\cite{Table}.
$V_{CS}$ can be equivalently
expressed as the sum of the Legendre functions:
\be
V_{CS}=1+
\frac{2 G M\nu}{\pi\sqrt{rr'}}\sum_{n=-\infty}^{\infty}
Q_{\nu |n|-1/2}(\cosh u_{0})\; e^{in\theta}\; .
\ee

In the limit of $\nu\rightarrow1$, $V_{CS}$ is reduced to
$V$ in the flat space~\cite{GHS,Shi1}.
On the other hand, in the limit of $r'\rightarrow0$,
$V_{CS}$ is reduced to
\be
V=1+\frac{2 G M\nu}
{\sqrt{r^{2}+(z-z')^{2}}}\; ,
\label{eq:solAFV}
\ee
which gives the metric representing
a cosmic string trapped by a black hole.
Thus the constant $M$ is identified to the black hole mass~\cite{AFV}.

The cosmic string-extreme black hole solution
in a general Einstein-Maxwell-dilaton model
with an arbitrary dilaton coupling~\cite{Shi1} can be constructed
in the same manner.


Next we will present the low-energy interaction in
the cosmic string-extreme black hole system.

The interaction energy of
the maximally charged dilatonic black holes in $R^{4}$
at low velocities has been calculated without a long-distance
approximation
by making use of the exact,
static solution~(\ref{eq:met}) with~(\ref{eq:sol})~\cite{Shi2,Shi3}.
Since there are only two-body velocity-dependent forces
in the multi-black hole system in this case,
the general expression for the interaction energy of $O(v^2)$
of an arbitrary number of extreme dilatonic black holes
can be easily obtained as:~\cite{Shi2,Shi3,Shi4}
\bea
H&=&\sum_{\alpha}\frac{1}{2}m_{\alpha}\Bv_{\alpha}^{2} +
\sum_{\alpha\beta}
\frac{G m_{\alpha}m_{\beta} |\Bv_{\alpha}-\Bv_{\beta}|^{2}}
{2|\Bx_{\alpha}-\Bx_{\beta}|}\nn
&=&\frac{1}{2}{\cal M}\BV^{2} +
\sum_{\alpha\beta}
\frac{m_{\alpha}m_{\beta} |\Bv_{\alpha}-\Bv_{\beta}|^{2}}{4 {\cal M}}
\left(1+
\frac{2 G {\cal M}}{|\Bx_{\alpha}-\Bx_{\beta}|}\right)\; ,
\label{eq:int}
\eea
where $\Bv_{\alpha}$ is the velocity of
the extreme dilatonic black hole with mass $m_{\alpha}$.
${\cal M}$ is the total mass, ${\cal M}=\sum_{\alpha}m_{\alpha}$.
$\BV$ is the velocity of
the center of mass; $\BV\equiv
{\sum_{\alpha} m_{\alpha} \Bv_{\alpha}}/{\cal M}$.


We use the method of images to obtain
the interaction energy of extreme dilatonic black holes.
We must note that the images of a certain black hole are
located at $\Lambda^{n}\Bx$ and have
the same mass $M$ and the velocity $\Lambda^{n}\Bv$.
We also note that the sum of the interaction energy must be
divided by the number of the images.
The number $p$ is extended to
a real number $\nu$ after completing the calculation.

Finally we get the following interaction energy up to $O(v^{2})$
of the extreme dilatonic black hole
with the mass $M$ and the velocity $\Bv$
on the conical space:%
\footnote{For a case of a general dilaton coupling,
since the interaction
contains many-body, velocity-dependent forces~\cite{Shi2,Shi3,Shi4},
the expression will be more complicated.}
\bea
H_{CS}&=&\frac{1}{2}Mv_{\parallel}^{2}+
\frac{1}{2}M
\left[1+\frac{2 G M}{\rho}
\tan\frac{\pi(\nu-1)}{2 \nu}\right]\Bv_{\perp}^{2}\nn
&=&\frac{1}{2}Mv_{\parallel}^{2}+
\frac{1}{2}M
\left[1+\frac{2 G M}{\rho}
\tan\frac{\Delta}{4}\right]\Bv_{\perp}^{2}\;,
\label{eq:intCS}
\eea
where  $\rho$ is the distance between the cosmic string
and the extreme black hole.
$v_{\parallel}$ is the $z$ component of the velocity
of the extreme dilatonic black hole, while $\Bv_{\perp}$
is the component of the velocity perpendicular to
the cosmic string.
It is quite reasonable that the interaction energy diverges
in the limit of $\Delta\rightarrow 2\pi$.

The characteristic length scale of the interaction
\be
\rho_{0}\equiv G M \tan\frac{\Delta}{4}\approx 2\pi G^{2}M\mu\; ,
\ee
may be very small, if the string parameter
$G\mu\simeq 10^{-6}\sim 10^{-5}$ as expected
from cosmological considerations~\cite{VS}.


We shall study the classical scattering
using the low-energy interaction energy.
We consider the scattering of a maximally charged
dilatonic black hole by a cosmic string
by using the metric of moduli space~\cite{Manton}.

Hereafter we assume that the black holes move
in a plane which is perpendicular to the string.
Therefore the moduli space of this configuration
is reduced to be a two-dimensional space
parameterized by the distance $\rho$,
the azimuthal angle $\varphi$.
For this two-body system, the metric on moduli space which
spanned by the coordinate $\xi$
can be read from Eq.~(\ref{eq:intCS}) as
\be
ds_{MS}^2=
\gamma_{ij}d\xi^{i}d\xi^{j}=
\gamma(\rho)
\left(d\rho^{2}+\frac{\rho^{2}}{\nu^{2}}d\varphi^{2}\right)\; ,
\label{eq:le1}
\ee
with
\be
\gamma(\rho)=1+\frac{2 G M}{\rho}\tan\frac{\Delta}{4}=
1+\frac{2\rho_{0}}{\rho}\; .
\label{eq:le2}
\ee
Here $\varphi$ stands for the azimuthal angle which has
a range $0\leq\varphi<2\pi$.

The path of the moving extreme black hole is determined by
the geodesic equation on the moduli space~\cite{Manton},
because the geodesic on the metric~(\ref{eq:le1},~\ref{eq:le2})
realizes the path of minimal energy.
We find that the scattering trajectory satisfies
the following differential equation:
\be
\left(\frac{du}{d\varphi}\right)^{2}+\nu^{2}u^{2}=
\frac{1}{b^{2}}\left(1+2\rho_{0} u\right)\; ,
\label{eq:de}
\ee
where $u=1/\rho$ and $b$ is the impact parameter.

The scattering angle $\Theta$ can be obtained by solving the
equation~(\ref{eq:de}) and expressed as:
\bea
\frac{\Theta}{\nu}&=&
\left(1-\frac{1}{\nu}\right)\pi+
2\arctan\left(\frac{\rho_{0}\nu}{b}\right)\nn
&=&\frac{\Delta}{2}+
2\arctan\left(\frac{\rho_{0}\nu}{b}\right)\; .
\label{eq:Rut}
\eea
The angle of deflection is written as
the sum of the contributions from the
deficit angle~\cite{VS,Vil}
and the Rutherford scattering~\cite{Shi3}.


The quantum mechanical approach to the study of
the scattering process
is also possible by using the moduli space metric~\cite{FerEar,Shi3}.
The Schr\"odinger equation on the moduli space,
which is spanned by the coordinates $(\rho,\varphi,Z)$, can be read
as~\cite{FerEar,Shi3}
\bea
i\hbar\frac{\partial}{\partial t}\Psi&=&
-\frac{\hbar^{2}}{2 M}
\left(
\frac{1}{\sqrt{\det\gamma_{ij}}}\partial_{k}
\sqrt{\det\gamma_{ij}}\gamma^{k\ell}\partial_{\ell}+
\frac{\partial^{2}}{\partial Z^{2}}
\right)\Psi\nn
&=&-\frac{\hbar^{2}}{2 M}
\left[
\frac{1}{\gamma(\rho)}
\left(\frac{1}{\rho}\frac{\partial}{\partial\rho}\rho
\frac{\partial}{\partial\rho}+
\frac{\nu^{2}}{\rho^{2}}
\frac{\partial^{2}}{\partial \varphi^{2}}
\right)+
\frac{\partial^{2}}{\partial Z^{2}}
\right]\Psi\; .
\eea
When we assume the wave function $\Psi$ can be decomposed as
\be
\Psi=\psi(\rho,\varphi)\; e^{-i\frac{E}{\hbar}t} e^{ikZ}\; ,
\ee
we get the following wave equation:
\be
\left[
\frac{1}{\rho}
\frac{\partial}{\partial\rho}
\rho
\frac{\partial}{\partial\rho}+
\frac{\nu^{2}}{\rho^{2}}
\frac{\partial^{2}}{\partial \varphi^{2}}+
q^{2}\left(1+\frac{2\rho_{0}}{\rho}\right)
\right]\psi=0\; ,
\label{eq:same}
\ee
where $q^{2}\equiv 2ME/\hbar^{2}-k^{2}$.
The equation~(\ref{eq:same})
 is the same as Eq.~(4.2) in the paper
on the Coulomb problem on a cone by
Gibbons, Ruiz and Vachaspati~\cite{GRV},
if we replace $\rho\rightarrow r$, $\varphi\rightarrow \nu\phi$,
$q\rightarrow k$ and $q^{2}\rho_{0}\rightarrow \mu K$.
Thus we do not repeat the calculations on the scattering problem here.
One can find that the differential cross-section diverges
in the forward direction, but the divergence is shifted due to
the presence of deficit angle of the spacetime~\cite{GRV}.
This nature agrees with that of the classical scattering.


To summarize, we have studied the interaction of a cosmic string and
maximally charged dilatonic
black hole in the low velocity limit.
The scattering of the black hole by a cosmic string
has been investigated.
The effect of the conical structure of the spacetime
appears in the velocity-dependent force
but may be very small for a ``realistic'' cosmic string.

We must note that our analyses are done in the
low-velocity limit. The terms of higher order in $v$
and the radiational reaction may become important if the
typical scales of the scattering process are very small.
Moreover, cosmic strings may begin vibrating when the black holes
approach, and then the nature of the interaction will be complicated.
The study of these corrections needs
additional approximation schemes,
or requires numerical calculations.



\end{document}